\documentclass[11pt]{article}
\usepackage{algorithm}
\usepackage{amssymb,amsmath,amscd,latexsym,epic,eepic}
\usepackage{fullpage}

\newtheorem{theorem}{Theorem}
\newtheorem{definition}{Definition}

\newtheorem{remark}{Remark}

\newtheorem{corollary}{Corollary}

\def\qed{\rule{0.4em}{1.4ex}}

\newcommand{\RandState}{\mathsf{X}} 
\newcommand{\randstate}{\mathsf{X}} 
\newcommand{\randact}{\Theta} 
\newcommand{\RandAct}{\Theta} 
\newcommand{\pat}{\omega}
\newcommand{\Paths}{\Omega}
\newcommand{\straa}{\sigma}
\newcommand{\Straa}{\Sigma}
\newcommand{\strab}{\pi}
\newcommand{\Strab}{\Pi}
\newcommand{\mov}{{\Gamma}}
\newcommand{\moves}{A}
\newcommand{\gamegraph}{G}
\newcommand{\trans}{\delta}
\newcommand{\distr}{{\mathcal{D}}}
\newcommand{\dist}{{\mathcal{D}}}
\newcommand{\Distr}{{\mathcal{D}}}

\newcommand{\size}{\mathsf{size}}

\newcommand{\Prb}{\mathrm{Pr}}
\newcommand{\Exp}{\mathrm{E}}

\newcommand{\cale}{{\mathcal{E}}}

\newcommand{\vare}{\varepsilon}
\newcommand{\len}{\mathsf{len}}

\newcommand{\supp}{\mathrm{Supp}}

\newcommand{\Integer}{\mathbb{Z}}
\newcommand{\reals}{\mathbb{R}}

\newcommand{\set}[1]{\{\: #1 \:\}}

\newcommand{\va}{v_1}
\newcommand{\vb}{v_2}

\newcommand{\vas}{v_1}
\newcommand{\vbs}{v_2}
\newcommand{\wh}{\widehat}

\makeatletter

\begingroup \catcode `|=0 \catcode `[= 1
\catcode`]=2 \catcode `\{=12 \catcode `\}=12
\catcode`\\=12 |gdef|@xcomment#1\end{comment}[|end[comment]]
|endgroup

\def\@comment{\let\do\@makeother \dospecials\catcode`\^^M=10\def\par{}}

\def\begincomment{\@comment\@xcomment}

\makeatother

\newcommand{\slopefrac}[2]{\leavevmode\kern.1em
  \raise .5ex\hbox{\the\scriptfont0 #1}\kern-.1em
  /\kern-.15em\lower .25ex\hbox{\the\scriptfont0 #2}}

\begin{title}
{Stochastic Limit-Average Games are in EXPTIME}
\end{title}
\author{
 Krishnendu Chatterjee$^\dag$%
 \qquad Rupak Majumdar$^\S$
 \qquad Thomas A.\ Henzinger$^{\dag,\ddag}$ \\ [5pt]%
\normalsize
$\strut^\dag$ EECS, University of California, Berkeley, USA\\
\normalsize
$\strut^\S$ CS, University of California, Los Angeles, USA \\
\normalsize
$\strut^\ddag$ EPFL, Switzerland \\
\normalsize
{\tt \{c\_krish,tah\}@eecs.berkeley.edu, rupak@cs.ucla.edu}
}

\date{}

\begin{document}
\maketitle

\begin{abstract}
The value of a finite-state two-player zero-sum stochastic game with
limit-average payoff can be approximated to within $\varepsilon$ in
time exponential in a polynomial in the size of the
game times polynomial in logarithmic in $\frac{1}{\varepsilon}$, for all
$\varepsilon>0$.
\end{abstract}


\section{Introduction}

A zero-sum stochastic game is a repeated game over a finite
state space, played by two players.
Each player has a non-empty set of actions available at every state,
and in each round, each player chooses an action
from the set of available actions at  the current state simultaneously
with and independent from the other player.
The transition function is \emph{probabilistic}, and the next state
is determined by a probability distribution depending on the current state
and the actions chosen by the players.
In each round,
player~1 gets (and player~2 loses) a reward depending on the current state and
the actions chosen by the players.
The players are informed of the history of the play consisting of
the sequence of states visited and the actions of the players played so
far in the play.
A strategy for a player is a recipe to extend the play: given a finite
sequence of states and pairs of actions
representing the history of the play, a strategy
specifies a probability distribution over the set of available actions
at the last state of the history.
The \emph{limit-average} for player~1 reward of a pair of strategies
$\straa$ and $\strab$ for player~1 and player~2, respectively, and a
starting state $s$ is defined as
\[
\va(s,\straa,\strab) = \Exp_s^{\straa,\strab} \lim\inf_{n\to \infty}
\bigg[\frac{1}{n} \cdot \sum_{i=1}^{n}
r(\RandState_i,\RandAct_{i,1},\RandAct_{i,2})\bigg],
\]
where  $\RandState_i$ is the random variable for the state reached at
round $i$ of the game, and $\RandAct_{i,j}$ is the random variable for the action played by
player~$j$ at round $i$ of the game, under strategies $\straa$ and $\strab$
and starting state $s$,
and $r(s,a,b)$ gives the reward at state $s$ for actions $a$ and $b$.
The form of the objective explains the term  limit average.
First the average is taken with respect to the expected rewards in the
first $n$ rounds of the game.
Then the objective is defined as the liminf of these averages.
A stochastic game with a limit-average reward is called a limit-average game.
The fundamental question in stochastic games is the existence of a {\em value}, that is,
whether
\[
\sup_{\straa \in \Straa} \inf_{\strab\in \Strab} \va(s,\straa,\strab) =
\inf_{\strab\in \Strab}\sup_{\straa\in \Straa} \va (s,\straa,\strab),
\]
where $\Straa$ and $\Strab$ denote the sets of strategies for player~1
and player~2, respectively.

Stochastic games were introduced by Shapley \cite{Shapley53}, who showed
the existence of a value in {\em discounted} games,
where the game stops at each round with probability $\beta$,
for some $0< \beta < 1$, and the goal of a player is to maximize the expectation 
of the total sum of the rewards.
Limit-average games were introduced by Gillette \cite{Gillette57}, who studied
the special cases of perfect information
(in each round, at most one player has a choice of moves)
and irreducible stochastic games.
The existence of a value for the perfect information case was proved in
\cite{LiggettLippman69}.
Gillette's paper also introduced a limit-average
game called the Big Match, which was solved in \cite{BlackwellFerguson68}.
Bewley and Kohlberg \cite{BK76} then showed how Pusieux series expansions
can be used for the asymptotic analysis of discounted games.
This, and the winning strategy in the Big Match, was used
by Mertens and Neyman \cite{MN81} to show the existence
of a value in limit-average games.

While the {\em existence} of a value in general limit-average
stochastic games has been extensively studied, the {\em computation}
of values has received less attention.\footnote{In this paper we 
take the classical view of computation, where an algorithm 
either answers ``Yes'' or ``NO'', or outputs a set of rational numbers} 
In general, it may happen
that a game with rational rewards and rational transition
probabilities still has an irrational value~\cite{RagFil91}. Hence,
we can only hope to have approximation algorithms that compute the
value of a game up to a given approximation $\vare$, for a real
$\vare>0$. Even the approximation of values is not simple, because
in general limit-average games only admit $\eta$-optimal strategies,
for all reals $\eta>0$, rather than optimal
strategies~\cite{BlackwellFerguson68}, and the $\eta$-optimal
strategies of~\cite{MN81} require
infinite memory. This precludes, for example, common algorithmic
techniques that enumerate over certain finite sets of strategies
and, having fixed a strategy, solve the resulting Markov decision
process using linear programming techniques~\cite{FV97}. Most
research has therefore characterized particular subclasses of games
for which \emph{stationary} optimal strategies exist (a stationary
strategy is independent of the history of a play and depends only on
the current state) \cite{ParthasarathyRaghavan81,HoffmanKarp} (see
\cite{FV97} for a survey), and the main algorithmic tool has been
value or policy iteration, which can be shown to terminate in an
exponential number of steps (but often behaves better in practice)
for many of these particular classes.

In this paper, we characterize the computational complexity of approximating the
value of a limit-average game.
We show that for any given real $\varepsilon > 0$,
the value of a game $G$ at a state
can be computed to within $\varepsilon$-precision
in time bounded by an exponential in a polynomial in the size of the
game $G$ times a polynomial function of $\log\frac{1}{\varepsilon}$.
This shows that approximating the value of limit-average games lies in the
computational complexity class EXPTIME \cite{Papa94}.
Our main technique is the characterization of values as semi-algebraic
quantities \cite{BK76,MN81}.
We show that for a real number $\alpha$, whether the value of a stochastic
limit-average game at a state $s$ is strictly greater than $\alpha$ can be
expressed as a sentence in the theory of real-closed fields.
Moreover, this sentence is polynomial in the
size of the game and has a constant number of quantifier alternations.
The theory of real-closed fields is decidable in time exponential in the size
of a formula and doubly exponential in the quantifier alternation
depth \cite{Basu99}.
This, together with binary search over the range of values,
gives an algorithm exponential in polynomial in the size of
the game graph times polynomial in logarithmic in $\frac{1}{\vare}$ to
approximate the value, for $\vare>0$.
Our techniques combine several known results to provide the first
complexity bound on the general problem of approximating the value of
stochastic games with limit-average objectives.
It may be noted that the best known deterministic algorithm for the special case of
perfect information limit-average games also requires exponential time.

\section{Definitions}

\noindent{\bf Probability distributions.}
For a finite set~$A$, a {\em probability distribution\/} on $A$ is a
function $\trans\!:A\to[0,1]$ such that $ \sum_{a \in A} \trans(a) = 1$.
We denote the set of probability distributions on $A$ by $\distr(A)$.
For a distribution $\trans \in \distr(A)$, we denote by
$\supp(\trans) = \{x \in A \mid \trans(x) > 0\}$ the {\em support\/}
of $\trans$.

\begin{definition}[Stochastic games]
A (two-player zero-sum) {\em stochastic game\/} $\gamegraph = \langle S,
\moves, \mov_1, \mov_2, \trans, r \rangle$ consists of the following components.

\begin{itemize}

\item A finite set $S$ of states.

\item A finite set $\moves$ of moves or actions.

\item Two move assignments $\mov_1, \mov_2 \!: S\to 2^\moves
    \setminus \emptyset$.  For $i \in \{1,2\}$, assignment
    $\mov_i$ associates with each state $s \in S$ a non-empty
    set $\mov_i(s) \subseteq \moves$ of moves available to player $i$
    at state $s$.

\item
A probabilistic transition function
$\trans: S \times \moves \times \moves \to \Distr(S)$ that gives the
probability $\trans(s, a, b)(t)$ of a transition from $s$ to
$t$ when player~1 plays move $a$ and player~2 plays move $b$,
for all $s,t\in S$  and $a \in \mov_1(s)$, $b \in \mov_2(s)$.

\item A reward function $r : S \times \moves \times \moves \to \reals$
that maps every state and pair of moves to a real-valued reward.
\qed
\end{itemize}
\end{definition}
The special class of \emph{perfect-information} games can be obtained
from stochastic games with the restriction that for all $s \in S$
either $|\mov_1(s)|=1$ or $|\mov_2(s)|=1$, i.e., at every state at most one
player can influence the transition.
If the transition function $\trans$ is deterministic rather than probabilistic
then we call the game a \emph{deterministic} game.
The class of \emph{rational} stochastic games are the special class
of stochastic games such that all rewards and transition probabilities
are rational.

\medskip\noindent{\bf Size of a stochastic game.}
Given a stochastic game $G$ we use the following notations:
\begin{enumerate}
\item $n=|S|$ is the number of states;
\item $|\trans|= \sum_{s \in S} |\mov_1(s)| \cdot |\mov_2(s)|$ is the
number of entries of the transition function.
\end{enumerate}
Given a rational stochastic game we use the following notations:
\begin{enumerate}
\item $\size(\trans)= \sum_{t \in S} \sum_{a \in \mov_1(s)}
\sum_{b \in \mov_2(s)} |\trans(s,a,b)(t)|$,
where $|\trans(s,a,b)(t)|$ denotes the space to express $\trans(s,a,b)(t)$ in
binary;
\item $\size(r)=\sum_{s \in S} \sum_{a \in \mov_1(s)}
\sum_{b \in \mov_2(s)} |r(s,a,b)|$, where $|r(s,a,b)|$ denotes the space to
express $r(s,a,b)$ in binary;

\item $|G|=\size(G)= \size(\trans) + \size(r)$.
\end{enumerate}
The specification of a game $G$ requires $O(|G|)$ bits.
Given a stochastic game with $n$ states, we assume without loss of generality
that the state space of the stochastic game structure is enumerated as natural numbers,
$S =\set{1,2,\ldots,n}$, i.e., the states are numbered from $1$ to $n$.

At every state $s\in S$, player~1 chooses a move $a\in\mov_1(s)$,
and simultaneously and independently player~2 chooses a move $b\in\mov_2(s)$.
The game then proceeds to the successor state $t$ with probability
$\trans(s,a,b)(t)$, for all $t \in S$.
At the state $s$, for moves $a$ for player~1 and $b$ for player~2,
player~1 wins and player~2 loses a reward of value $r(s,a,b)$.

A {\em path\/} or a \emph{play} $\pat$ of $\gamegraph$ is an infinite sequence
$\pat = \langle s_0, (a_0,b_0),s_1,(a_1,b_1), s_2, (a_2,b_2),\ldots \rangle $
of states and pairs of moves such that $(a_i,b_i) \in
\mov_1(s_i) \times \mov_2(s_i)$  and $s_{i+1} \in \supp(\trans(s_i,a_i,b_i))$,
for all $i\geq 0$.
We denote by $\Paths$ the set of all paths, and by $\Paths_s$ the set of all
paths starting from state $s$.

\medskip\noindent
{\bf Randomized strategies.}
A \emph{strategy} for player~1 is a function
$\straa:  (S \times \moves\times \moves)^* \cdot S \to \distr(A)$
that associates with every prefix of a play, representing the history
of the play so far, and the current state a probability distribution from
$\Distr(A)$ such that for all $w \in (S \times \moves \times \moves)^*$ and
all $s \in S$, we have $\supp(\straa(w \cdot s)) \subseteq \mov_1(s)$.
Observe that the strategies can be randomized (i.e., not
necessarily deterministic) and history-dependent
(i.e., not necessarily stationary).
Similarly we define strategies $\strab$ for player~2.
We denote by $\Straa$ and $\Strab$ the sets of strategies for
player $1$ and player $2$, respectively.

Once the starting state $s$ and the strategies $\straa$ and $\strab$
for the two players have been chosen, the game is reduced to a
stochastic  process.
Hence, the probabilities of events are uniquely defined, where an {\em
event\/} $\cale\subseteq\Paths_s$ is a measurable set of
paths.
For an event $\cale\subseteq\Paths_s$, we denote by
$\Prb_s^{\straa,\strab}(\cale)$ the probability that a path belongs to
$\cale$ when the game starts {f}rom $s$ and the players follow the
strategies $\straa$ and~$\strab$.
We denote by $\Exp_s^{\straa,\strab}[\cdot]$ the associated expectation
operator with the probability measure $\Prb_s^{\straa,\strab}(\cdot)$.
For $i \geq 0$, we  denote by $\RandState_i: \Paths \to S$
the random variable denoting the $i$-th state along a path, and for
$j \in \set{1,2}$, we denote by
$\RandAct_{i,j}: \Paths_s \to \moves$ the
random variable denoting the move of player~$j$ in the $i$-th round of a
play.

\medskip\noindent
{\bf Limit-average payoff.}
Let $\straa$ and $\strab$ be strategies of player~1 and player~2, respectively.
The {\em limit-average} payoff $\va(s,\straa,\strab)$ for player~1 at a state $s$,
for the strategies $\straa$ and $\strab$, is defined as
\[
\va(s,\straa,\strab) = \Exp_s^{\straa,\strab}
\lim\inf_{N \to \infty} \bigg[ \frac{1}{N} \cdot \sum_{i=1}^{N}
r(\RandState_i,\RandAct_{i,1},\RandAct_{i,2})\bigg].
\]
Similarly, for player~2, the payoff $\vb(s,\straa,\strab)$ is defined as
\[
\vb(s,\straa,\strab) =
\Exp_s^{\straa,\strab} \lim\sup_{N \to \infty} \bigg[\frac{1}{N} \cdot \sum_{i=1}^{N}
-r(\RandState_i,\RandAct_{i,1},\RandAct_{i,2})\bigg].
\]
In other words,
player~1 wins and player~2 looses the ``long-run'' average of the
rewards of the play.
A stochastic game $G$ with limit-average payoff is called a stochastic
limit-average game.

Given a state $s\in S$ and we are interested in finding the maximal payoff
that player $1$ can ensure against all strategies for player~2,
and the maximal payoff that player~2 can ensure against all strategies for
player~1.
We call such payoff the {\em value} of the game $\gamegraph$ at $s$ for
player $i\in\set{1,2}$.
The values for player~1 and player~2 are
defined for all $s \in S$ by
\[
\begin{array}{lcl}
  \vas(s) =
  \sup_{\straa\in\Straa}\inf_{\strab\in\Strab}
  \va(s,\straa,\strab)
& \mbox{and} &
  \vbs(s) =
  \sup_{\strab\in\Strab}\inf_{\straa\in\Straa}
  \vb(s,\straa,\strab).
\end{array}
\]
Mertens and Neyman~\cite{MN81} established the determinacy of stochastic
limit-average games.

\begin{theorem}{\bf \cite{MN81}}
For all stochastic limit-average games $\gamegraph$ and for all states $s$ of $G$,
we have $\vas(s) + \vbs(s)=0$.
\end{theorem}

\medskip\noindent{\bf Stronger notion of existence of values~\cite{MN81}.}
The values for stochastic limit-average games exist in a strong
sense:
for all reals $\vare>0$, there exist strategies $\straa^* \in \Straa, \strab^* \in \Strab$
such that the following conditions hold:
\begin{enumerate}
\item for all $\straa \in \Straa$ and $\strab \in \Strab$, we have
\begin{eqnarray}\label{eq-strong-value1}
-\vare +
\Exp_s^{\straa,\strab^*} \lim\sup_{N \to \infty}
\bigg[\frac{1}{N} \cdot \sum_{i=1}^{N} r(\RandState_i,\RandAct_{i,1},\RandAct_{i,2})\bigg]
\leq
\Exp_s^{\straa^*,\strab} \lim\inf_{N \to \infty}
\bigg[\frac{1}{N} \cdot \sum_{i=1}^{N} r(\RandState_i,\RandAct_{i,1},\RandAct_{i,2})\bigg]
 + \vare;
\end{eqnarray}

\item there exists an integer $N_0$ such that for all
$\straa \in \Straa$ and $\strab \in \Strab$, for all integers $N \geq N_0$, we have
\begin{eqnarray}
-\vare + \Exp_s^{\straa,\strab^*}\bigg[\frac{1}{N} \cdot
\sum_{i=1}^{N} r(\RandState_i,\RandAct_{i,1},\RandAct_{i,2})\bigg]
\leq \Exp_s^{\straa^*,\strab}\bigg[\frac{1}{N} \cdot \sum_{i=1}^{N}
r(\RandState_i,\RandAct_{i,1},\RandAct_{i,2})\bigg] + \vare.
\end{eqnarray}

\item there exists  $0<\beta_0<1$ such that for all
$\straa \in \Straa$ and $\strab \in \Strab$, for all  $0<\beta \leq
\beta_0$, we have
\begin{eqnarray}
-\vare + \Exp_s^{\straa,\strab^*}\bigg[\beta \cdot
\sum_{i=1}^{\infty}
(1-\beta)^{i-1}r(\RandState_i,\RandAct_{i,1},\RandAct_{i,2})\bigg]
\leq \Exp_s^{\straa^*,\strab}\bigg[\beta \cdot \sum_{i=1}^{\infty}
(1-\beta)^{i-1} r(\RandState_i,\RandAct_{i,1},\RandAct_{i,2})\bigg]
+ \vare.
\end{eqnarray}

\end{enumerate}
Let
$\overline{v}_1(s,\straa,\strab)
=\displaystyle\Exp_s^{\straa,\strab} \lim\sup_{N \to \infty}
\bigg[\frac{1}{N}\cdot \sum_{i=1}^{N} r(\RandState_i,\RandAct_{i,1},\RandAct_{i,2})\bigg]$,
then (\ref{eq-strong-value1}) is equivalent to the following equality:
\[
\sup_{\straa \in \Straa} \inf_{\strab \in \Strab} \va(s,\straa,\strab)
=\inf_{\strab \in \Strab} \sup_{\straa \in \Straa}
\overline{v}_1(s,\straa,\strab).
\]

\newcommand{\abs}{\mathsf{abs}}
\section{Theory of Real-closed Fields and Quantifier Elimination}

Our main technique is to represent the
value of a game as a formula in the theory of real-closed fields.
We denote by ${\mathbf R}$ the real-closed field $(\reals, +,\cdot,0,1,\leq)$
of the reals with addition and multiplication.
In the sequel we write ``real-closed field''
to denote the real-closed field  ${\mathbf R}$.
An {\em atomic formula} is an expression of the form $p < 0$ or
$p=0$, where
$p$ is a (possibly) multi-variate polynomial with coefficients in the
real-closed field.
Coefficients are rationals or symbolic constants (e.g., the symbolic constant
$e$ stands for $2.71828\ldots$).
We will consider the special case when only rational
coefficients of the form $\frac{q_1}{q_2}$, where $q_1,q_2$ are integers, are allowed.
A {\em formula} is constructed from atomic formulas by the grammar
\[
\varphi ::= a \mid \neg a \mid \varphi \land \varphi \mid \varphi \lor \varphi
\mid \exists x. \varphi \mid \forall x. \varphi,
\]
where $a$ is an atomic formula, $\neg a$ denotes complement of $a$,
$\varphi_1 \land \varphi_2$ denotes conjunction of $\varphi_1$ and
$\varphi_2$, $\varphi_1\lor \varphi_2$ denotes disjunction of
$\varphi_1$ and $\varphi_2$,
and $\exists$ and $\forall$ denote existential
and universal quantification, respectively.
We use the standard abbreviations such as $p\leq 0,p\geq 0$ and $p>0$ that
are derived as follows:
\[
\ p \leq 0 \  \text{  (for $p < 0 \lor p = 0$),} \qquad
\ p \geq 0 \  \text{  (for $\neg (p < 0)$),} \qquad
\mbox{and} \quad \ p > 0 \ \text{ (for $\neg (p \leq 0)$)}.
\]
The semantics of formulas are given in a standard way.
A variable $x$ is {\em free} in the formula $\varphi$ if it is not in the scope of
a quantifier $\exists x$ or $\forall x$.
A {\em sentence} is a formula with no free variables.
A formula is {\em quantifier-free} if it does not contain any existential
or universal quantifier.
Two formulas $\varphi_1$ and $\varphi_2$ are \emph{equivalent} if the set of free variables
of $\varphi_1$ and $\varphi_2$ are the same, and for every assignment
to the free variables the formula $\varphi_1$ is true if and only if
the formula $\varphi_2$ is true.
A formula $\varphi$ admits {\em quantifier elimination} if there is an
algorithm  to convert it to an equivalent quantifier-free formula.
A quantifier elimination algorithm takes as input a formula $\varphi$
and returns an equivalent quantifier-free formula, if one exists.

Tarski proved that
every formula in the theory of real-closed fields admits quantifier elimination,
and (by way of quantifier elimination) that there is an
algorithm to decide the truth of a sentence $\varphi$ in the
theory of real-closed fields
(see~\cite{Tarski51} for algorithms that decide the truth of a 
sentence $\varphi$ in the theory of real-closed fields).
The complexity of the algorithm of Tarski has subsequently improved,
and we now present a result of Basu~\cite{Basu99} on the complexity of
quantifier elimination for formulas in the theory of the real-closed field.

\medskip\noindent{\bf Complexity of quantifier elimination.}
We first define the length of a formula $\varphi$, and then define the size of
a formula with rational coefficients.
We denote the length and size of $\varphi$ as $\len(\varphi)$ and
$\size(\varphi)$, respectively.
The length of a polynomial $p$ is defined as the sum of the
length of its constituent monomials plus the number of monomials in the polynomial.
The length of a monomial is defined as its degree plus the number of
variables plus 1 (for the coefficient).
For example, for the monomial $\frac{1}{4} \cdot x^3 \cdot y^2 \cdot z$, its
length is $6 + 3+1=10$.
Given a polynomial $p$, the length of both $p < 0$ and $p=0$ is $\len(p) + 2$.
This defines the length of an atomic formula $a$.
The length of a formula $\varphi$ is inductively defined as follows:
\[
\begin{array}{rcl}
\len(\neg a) & = & \len(a) +1; \\
\len(\varphi_1 \land \varphi_2) & = & \len(\varphi_1) + \len(\varphi_2) +1; \\
\len(\varphi_1 \lor \varphi_2) & = & \len(\varphi_1) + \len(\varphi_2) +1; \\
\len(\exists x. \varphi) & = & \len(\varphi) +2; \\
\len(\forall x. \varphi) & = & \len(\varphi) +2. \\
\end{array}
\]
Observe that the length of a formula is defined for formulas that
may contain symbolic constants as coefficients.
For formulas with rational coefficients we define its size as follows:
the size of $\varphi$, i.e., $\size(\varphi)$, is defined as the sum of $\len(\varphi)$ and
the space required to specify the rational coefficients of the
polynomials appearing in $\varphi$ in binary.
We state a result of Basu~\cite{Basu99} on the complexity of quantifier
elimination for the real-closed field.
The following theorem is a specialization of Theorem~1 of~\cite{Basu99};
also see Theorem 14.14 and Theorem 14.16 of~\cite{BasuBook}.

\begin{theorem} \label{thrm:basu}
{\bf \cite{Basu99}}
Let $d,k,m$ be nonnegative integers, $X=\set{X_1,X_2,\ldots, X_k}$ be a
set of $k$ variables, and
${\cal P}=\set{p_1,p_2,\ldots,p_m}$ be a set of $m$ polynomials
over the set $X$ of variables,
each of degree at most $d$ and with coefficients in the real-closed field.
Let $X_{[r]}, X_{[r-1]},\ldots,X_{[1]}$ denote a partition of the
set $X$ of variables into $r$ subsets such that the set $X_{[i]}$ of variables
has size $k_i$, i.e., $k_i =|X_{[i]}|$ and $\sum_{i=1}^r k_i=k$.
Let
\[
\Phi = (Q_r X_{[r]}). \ (Q_{r-1} X_{[r-1]}). \ \cdots . (Q_2 X_{[2]}). \ (Q_1 X_{[1]}). \
\varphi(p_1,p_2,\ldots,p_m)
\]
be a sentence with $r$ alternating quantifiers
$Q_i \in \set{\exists,\forall}$ (i.e., $Q_{i+1} \neq Q_i$),
and
$\varphi(p_1,p_2,\ldots,p_m)$ is a quantifier-free formula
with atomic formulas of the form $p_i \bowtie 0$, where $\bowtie \ \in \set{<,>,=}$.
Let $D$ denote the ring generated by the coefficients of the polynomials
in ${\cal P}$.
Then the following assertions hold.
\begin{enumerate}
\item There is an algorithm to decide the truth of $\Phi$ using
\[
m^{\prod_i (k_i+1)} \cdot d ^{\prod_i O(k_i)} \cdot \len(\varphi)
\]
arithmetic operations (multiplication, addition, and sign determination) in
$D$.

\item If $D=\Integer$ (the set of integers) and the bit sizes of the
coefficients of the polynomials are bounded by $\gamma$, then the bit sizes of the
integers appearing in the intermediate computations of the truth of
$\Phi$ is bounded by
\[
\gamma \cdot d^{\prod_i O(k_i)}.
\]
\end{enumerate}
\end{theorem}

The result of part~1 of Theorem~\ref{thrm:basu}
holds for sentences with symbolic constants as coefficients.
The result of part~2 of Theorem~\ref{thrm:basu}
is for the special case of sentences with only integer coefficients.
Part~2 of Theorem~\ref{thrm:basu} follows from the results
of~\cite{Basu99}, but is not explicitly stated as a theorem there;
for an explicit statement as a theorem, see Theorem~14.14 and
Theorem~14.16 of~\cite{BasuBook}.

\begin{remark}\label{remark:lim-avg-integer}
Given two integers $a$ and $b$, let $|a|$ and $|b|$ denote the space to express
$a$ and $b$ in binary, respectively.
The following assertions hold: given integers $a$ and $b$,
\begin{enumerate}
\item given signs of $a$ and $b$, the sign determination of $a+b$ can be done
in $O(|a|+|b|)$ time, i.e., in linear time, and
the sign determination of $a\cdot b$ can be done $O(1)$ time, i.e., in constant time;
\item addition of $a$ and $b$ can be done in $O(|a| + |b|)$ time, i.e., in linear time; and
\item multiplication of $a$ and $b$ can be done in $O(|a| \cdot |b|)$ time, i.e., in
quadratic time.
\end{enumerate}
It follows from the above observations, along with Theorem~\ref{thrm:basu},
that if $D=\Integer$ and the bit sizes of the coefficients of the polynomials
appearing in $\Phi$ are bounded by $\gamma$,
then the truth of $\Phi$ can be determined in time
\begin{eqnarray}\label{eq-complexity1}
m^{\prod_i O(k_i+1)} \cdot d^{\prod_i O(k_i)} \cdot O(\len(\varphi) \cdot \gamma^2).
\end{eqnarray}
\end{remark}

\section{Computation of Values in Stochastic Games}

The values in stochastic limit-average games  can
be irrational even if all  rewards and  transition probability values
are rational~\cite{RagFil91}.
Hence, we can algorithmically only approximate the values to within a
precision $\vare$, for $\vare>0$.

\medskip\noindent{\bf Discounted value functions.}
Let $G$ be a stochastic game with reward function $r$.
For a real $\beta$, with $0<\beta<1$,
the $\beta$-discounted value function $\va^\beta$ is defined as
follows:
\[
\va^\beta(s)=\sup_{\straa \in \Straa} \inf_{\strab \in \Strab} \ \
\beta \cdot \Exp_{s}^{\straa,\strab}\big[\sum_{i=1}^\infty (1-\beta)^i \cdot
r(\randstate_i,\randact_{i,1},\randact_{i,2})\big].
\]
For a stochastic game $G$,
the $\beta$-discounted value function $\va^\beta$ is monotonic with respect
to $\beta$ in a neighborhood of $0$~\cite{BK76}.

\subsection{Sentence for the value of a stochastic game}
We now describe how we can obtain a sentence
in the theory of the real-closed field that states that the value of a stochastic
limit-average game at a given state is
strictly greater than $\alpha$, for a real $\alpha$.
The sentence applies to the case where the rewards and the transition
probabilities are specified as symbolic or rational constants.

\medskip\noindent{\bf Formula for $\beta$-discounted value functions.}
Given a real $\alpha$ and a stochastic limit-average game $G$,
we present a formula in the theory of the real-closed
field to express that the $\beta$-discounted value $\va^\beta(s)$
at a given state $s$ is strictly greater than $\alpha$, for $0<\beta<1$.
A {\em valuation\/} $v\in \reals^n$ is a vector of reals, and for $1\leq i \leq n$,
the $i$-th component of $v$ represents the value $v(i)$ for state $i$.
For every state $s \in S$ and for every move $b \in \mov_2(s)$
we define a polynomial $u_{(s,b,1)}$ for player~1 as a function of
$x \in \dist(\mov_1(s))$, a valuation $v$ and $0<\beta < 1$ as follows:
\[
u_{(s,b,1)}(x,v,\beta) = \beta \cdot \sum_{a \in \mov_1(s)} x(a) \cdot r(s,a,b) \ + \
    (1-\beta) \cdot \sum_{a \in \mov_1(s)} x(a) \cdot \sum_{t \in S}
    \trans(s,a,b)(t) \cdot v(t) -v(s).
\]
The polynomial $u_{(s,b,1)}$ consists of the variables
$\beta$, and $x(a)$ for $a \in \mov_1(s)$,
and $v(t)$ for  $t \in S$.
Observe that given a stochastic limit-average game,
$r(s,a,b)$ for $a \in \mov_1(s)$, and $\trans(s,a,b)(t)$ for $t\in S$ and
$a \in \mov_1(s)$ are rational or symbolic constants given by the game graph,
not variables.
The coefficients of the polynomial are $r(s,a,b)$ for $a \in \mov_1(s)$, and
$\trans(s,a,b)(t)$ for $a \in \mov_1(s)$ and $t \in S$.
Hence the polynomial has degree~$3$ and has $1+ |\mov_1(s)| + n$ variables.
Similarly, for $s\in S$, $a \in \mov_1(s)$, $y \in \dist(\mov_2(s))$,
$v \in \reals^n$, and $0<\beta<1$, we have polynomials $u_{(s,a,2)}$  defined by
\[
u_{(s,a,2)}(y,v,\beta) = \beta \cdot \sum_{b \in \mov_2(s)} y(b)\cdot r(s,a,b) +
    (1-\beta) \cdot \sum_{b \in \mov_2(s)} y(b) \cdot \sum_{t \in S}
    \trans(s,a,b)(t) \cdot v(t) -v(s).
\]
The sentence stating that $\va^\beta(s)$ is strictly greater than
$\alpha$ is as follows.
We have variables $x_{s}(a)$ for $s \in S$ and $a \in \mov_1(s)$,
$y_{s}(b)$ for $s \in S$ and $b\in \mov_2(s)$, and
variables $v(1), v(2),\ldots, v(n)$.
For simplicity we write
$x_s$ for the vector of variables $x_s(a_1), x_s(a_2),\ldots, x_s(a_j)$,
where $\mov_1(s)=\set{a_1,a_2,\ldots,a_j}$,
$y_s$ for the vector of variables $y_s(b_1), y_s(b_2),\ldots, y_s(b_l)$,
where $\mov_2(s)=\set{b_1,b_2,\ldots,b_l}$,
and $v$ for the vector of variables $v(1), v(2), \ldots, v(n)$.
The sentence is as follows:
\[
\begin{array}{rclclcl}
\Phi_\beta (s,\alpha) & = &
\exists x_{1},\ldots, x_{n}. \ \exists y_{1},\ldots, y_{n}. \ \exists v.
& & \Psi(x_1,x_2,\ldots,x_n,y_1,y_2,\ldots,y_n)
& &  \\[2ex]
& \bigwedge &
\displaystyle
\bigwedge_{s \in S, b \in \mov_2(s)} \big(u_{(s,b,1)}(x_s,v,\beta) \geq 0 \big)
& \bigwedge &
\displaystyle
\bigwedge_{s \in S, a \in \mov_1(s)} \big(u_{(s,a,2)}(y_s,v,\beta) \leq 0 \big)
& & \\[3ex]
& \bigwedge & (v(s) -\alpha >0); & & & &
\end{array}
\]
where $\Psi(x_1,x_2,\ldots,x_n, y_1,y_2,\ldots,y_n)$ specify the constraints that
$x_1,x_2,\ldots,x_n$ and $y_1,y_2,\ldots,y_n$ are valid randomized strategies and
is defined as follows:
\[
\begin{array}{rcl}
\Psi(x_1,x_2,\ldots,x_n, y_1,y_2,\ldots,y_n) &= &
\displaystyle
\bigwedge_{s \in S} \big( (\sum_{a \in \mov_1(s)} x_{s}(a))  -1=0\big)
 \ \wedge \
\bigwedge_{s\in S,a \in \mov_1(s)} \big(x_{s}(a) \geq 0\big)
   \\[2ex]
& \wedge  &
\displaystyle
\bigwedge_{s \in S} \big( (\sum_{b \in \mov_2(s)} y_{s}(b))  -1=0\big)
 \ \wedge \
\bigwedge_{s\in S,b \in \mov_2(s)} \big(y_{s}(b) \geq 0\big).
\end{array}
\]
The total number of polynomials in  $\Phi_\beta(s,\alpha)$ is
$1+\sum_{s \in S} (3\cdot|\mov_1(s)| + 3\cdot|\mov_2(s)| + 2)= O(|\trans|)$.
In the above formula we treat $\beta$ as a variable; it is a free
variable in $\Phi_\beta(s,\alpha)$.
Given a stochastic limit-average game $G$, for all $0 < \beta < 1$,
the correctness of $\Phi_\beta(s,\alpha)$ to specify that $\va^\beta(s) > \alpha$
can be proved from the results of~\cite{Shapley53}.

\medskip\noindent{\bf Value of a game as limit of discounted games.}
The result of Mertens-Neyman~\cite{MN81} established that the value
of a stochastic limit-average game is the limit of the $\beta$-discounted
values, as $\beta$ goes to~0.
Formally, we have
\[
\va(s) =\lim_{\beta \to 0^+} \va^\beta(s).
\]

\noindent{\bf Sentence for the value of a stochastic game.}
From the characterization of the value of a stochastic limit-average game
as the limit of the $\beta$-discounted values and the monotonicity
property of the $\beta$-discounted values in a neighborhood of $0$, we
obtain the following sentence $\Phi(s,\alpha)$ stating that the value at state~$s$ is
strictly greater than $\alpha$.
In addition to variables for $\Phi_\beta(s,\alpha)$, we  have
the variables $\beta$ and $\beta_1$.
The sentence $\Phi(s,\alpha)$ specifies the expression
\[
\exists \beta_1 >0.\  \forall \beta\in (0,\beta_1). \  \Phi_\beta(s,\alpha),
\]
and is defined as follows:
\[
\begin{array}{rclllcl}
\Phi(s,\alpha) & = & \exists \beta_1. \ \forall \beta. \
\exists x_1,\ldots, x_n. \ \exists y_1,\ldots, y_n. \ \exists v.&  &
\Psi(x_1,x_2,\ldots,x_n, y_1,y_2,\ldots,y_n)
& &
\\[2ex]
& \bigwedge &
(\beta_1 >0)  \bigwedge
\bigg[
(\beta_1 -\beta \leq 0)  \ \bigvee \ (\beta \leq 0) & \bigvee &
\bigg( (\beta_1 -\beta >0)  & & \\[2ex]
& & & & \bigwedge
\displaystyle
\bigwedge_{s \in S, b \in \mov_2(s)} \big( u_{(s,b,1)}(x_s,v,\beta) \geq 0 \big)\\[2ex]
& & & &\bigwedge
\displaystyle
\bigwedge_{s \in S, a \in \mov_1(s)} \big( u_{(s,a,2)}(y_s,v,\beta) \leq 0 \big)
\bigg) \bigg]
&  & \\[3ex]
& \bigwedge & (v(s) -\alpha >0); & & & &
\end{array}
\]
where $\Psi(x_1,x_2,\ldots,x_n, y_1,y_2,\ldots,y_n)$ specify the constraints that
$x_1,x_2,\ldots,x_n$ and $y_1,y_2,\ldots,y_n$ are  valid randomized strategies
(the same formula used for $\Phi_{\beta}(s,\alpha)$).\footnote{Our detailed formulas
$\Phi_{\beta}(s,\alpha)$ and $\Phi(s,\alpha)$ can be shortened, however, the present
formulas make it easier to understand the bound on parameters required for complexity bounds.}
Observe that $\Phi(s,\alpha)$ contains no free variable (i.e., the variables $x_s$,
$y_s$, $v$, $\beta_1$, and $\beta$ are quantified).
A similar sentence was used in~\cite{BK76} for values of discounted games.
The total number of polynomials in $\Phi(s,\alpha)$ is $O(|\trans|)$;
in addition to the $O(|\trans|)$ polynomials of $\Phi_\beta(s,\alpha)$ there are
$4$ more polynomials in $\Phi(s,\alpha)$.
In the setting of Theorem~\ref{thrm:basu} we obtain the following bounds
for $\Phi(s,\alpha)$:
\begin{eqnarray}\label{eq-bound1}
m=O(|\trans|);
\qquad k=O(|\trans|);
\qquad \prod_i (k_i+1)=O(|\trans|);
\qquad r=O(1);
\qquad d=3;
\end{eqnarray}
and hence we have
\[
m^{\prod_i (k_i+1)} \cdot d^{\prod_i O(k_i)}
=O(|\trans|)^{O(|\trans|)}
=2^{O\big(|\trans|\cdot \log(|\trans|)\big)}.
\]
Also observe that for a stochastic game $G$,
the sum of the lengths of the polynomials appearing in the
sentence is $O(|\trans|)$.
The present analysis along with Theorem~\ref{thrm:basu} yields
Theorem~\ref{thrm:value-qf-formula}.
The result of Theorem~\ref{thrm:value-qf-formula} holds for stochastic
limit-average games where the transition probabilities and
rewards are specified as symbolic constants.

\begin{theorem}\label{thrm:value-qf-formula}
Given a stochastic limit-average game $G$ with reward function $r$,
a state $s$ of $G$, and a real $\alpha$,
there is an algorithm to decide whether $\va(s)> \alpha$ using
$2^{O\big(|\trans| \cdot \log(|\trans|)\big)} \cdot O(|\trans|)$
arithmetic operations (addition, multiplication, and sign determination) in
the ring generated by the set
\[
\set{r(s,a,b) \mid s \in S, a \in \mov_1(s), b \in \mov_2(s)} \cup
\set{\trans(s,a,b)(t) \mid s,t \in S, a \in \mov_1(s), b \in \mov_2(s)}
\cup \set{\alpha}.
\]
\end{theorem}

\subsection{Algorithmic analysis}
For algorithmic analysis we consider rational stochastic games,
i.e., stochastic games such that $r(s,a,b)$ and $\trans(s,a,b)(t)$ are rational
for all states $s,t \in S$, and moves $a \in \mov_1(s)$ and $b\in \mov_2(s)$.
In the sequel we will only consider rational stochastic games.
Given the sentence $\Phi(s,\alpha)$ to specify that $\va(s) > \alpha$, we first reduce
it to an equivalent sentence $\wh{\Phi}(s,\alpha)$ as follows.
\begin{itemize}
\item For every rational coefficient  $\ell=\frac{q_1}{q_2}$, where $q_1,q_2 \in \Integer$, appearing in
$\Phi(s,\alpha)$ we apply the following procedure:
\begin{enumerate}
\item introduce a new variable $z_\ell$;
\item replace $\ell$ by $z_\ell$ in $\Phi(s,\alpha)$;
\item add a polynomial $q_2\cdot z_\ell-q_1=0$ as a conjunct to the quantifier-free body of the formula; and
\item existentially
quantify $z_\ell$ in the block of existential quantifiers after quantifying
$\beta_1$ and $\beta$.
\end{enumerate}
\end{itemize}
Thus we add $O(|\trans|)$ variables and polynomials, and increase the degree of
the polynomials in $\Phi(s,\alpha)$ by~1.
Also observe that the coefficients in $\wh{\Phi}(s,\alpha)$ are integers, and hence
the ring $\wh{D}$ generated by the coefficients in
$\wh{\Phi}(s,\alpha)$ is $\Integer$.
Similar to the bounds obtained in (\ref{eq-bound1}),
in the setting of Theorem~\ref{thrm:basu} we obtain the following bounds for
$\wh{\Phi}(s,\alpha)$:
\[
\wh{m}=O(|\trans|);
\qquad \wh{k}=O( |\trans|);
\qquad \prod_i (\wh{k}_i+1)=O(|\trans|);
\qquad \wh{r}=O(1);
\qquad \wh{d}=4;
\]
and hence
\[
\wh{m}^{\ \prod_i O(\wh{k}_i+1)\ }\cdot \wh{d}^{\ \prod_i O(\wh{k}_i)}
=O(|\trans|)^{O(|\trans|)}
=2^{O\big(|\trans|\cdot \log(|\trans|)\big)}.
\]
Also observe that the length of the sentence $\wh{\Phi}(s,\alpha)$ can be bounded
by $O(|\trans|)$, and the sum of the
bit sizes of the coefficients in $\wh{\Phi}(s,\alpha)$ can be bounded by
$O(|G| + |\alpha|)$,
where $|\alpha|$ is the space required to express $\alpha$ in binary.
This along with (\ref{eq-complexity1}) of Remark~\ref{remark:lim-avg-integer}
yields the following result.

\begin{theorem}\label{thrm:complexity}
Given a rational stochastic limit-average game $G$, a state $s$ of $G$, and
a rational $\alpha$, there is an algorithm that decides whether
$\va(s) > \alpha$ in time
\[
2^{O\big(|\trans|\cdot \log(|\trans|)\big)} \cdot O(|\trans|) \cdot
O(|G|^2 +|\alpha|^2)
=2^{O\big(|\trans|\cdot \log(|\trans|)\big)}  \cdot O(|G|^2 +|\alpha|^2).
\]
\end{theorem}

\subsection{Approximating the value of a stochastic game}
We now present an algorithm that approximates the value $\va(s)$
within a tolerance of $\vare>0$.
The algorithm (Algorithm~\ref{algo:conc-lim-avg})  is obtained
by a binary search technique along with the result of
Theorem~\ref{thrm:complexity}.
Algorithm~\ref{algo:conc-lim-avg} works for the special case of
\emph{normalized} rational stochastic games.
We first define normalized rational stochastic games and then present a
reduction of rational stochastic games to normalized rational stochastic
games.

\medskip\noindent{\bf Normalized rational stochastic games.}
A rational stochastic game is \emph{normalized} if the reward function satisfies
the following two conditions:
(1)~$\min\set{r(s,a,b) \mid s \in S, a \in \mov_1(s), b \in \mov_2(s)} \geq 0$; and
(2)~$\max\set{r(s,a,b) \mid s \in S, a \in \mov_1(s), b \in \mov_2(s)}\leq 1$.

\medskip\noindent{\bf Reduction.} We now present a reduction of rational
stochastic games to normalized rational stochastic games, such that
by approximating the values of normalized rational stochastic games we can
approximate the values of  rational stochastic games.
Given a reward function $r:S\times \moves \times \moves \to \reals$,
let
\[
M=\max\set{\abs(r(s,a,b)) \mid s \in S, a \in \mov_1(s), b \in \mov_2(s)},
\]
where $\abs(r(s,a,b))$ denotes the absolute value of $r(s,a,b)$.
Without loss of generality we assume $M>0$.
Otherwise, $r(s,a,b)=0$ for all states $s \in S$, and moves
$a \in \mov_1(s)$ and $b \in \mov_2(s)$, and hence $\va(s)=0$ for all
states $s \in S$ (i.e., the value function can be trivially computed).
Consider the reward function $r^+:S \times \moves \times \moves \to [0,1]$
defined as follows: for $s \in S$,  $a \in \mov_1(s)$, and $b \in \mov_2(s)$,
we have
\[
r^+(s,a,b)= \frac{r(s,a,b)+M}{2M}.
\]
The reward function $r^+$ is normalized and the following assertion hold.
Let $\va$ and $\va^+$ denote the value functions for the reward functions
$r$ and $r^+$, respectively.
Then for all states $s\in S$ we have
\[
\va^+(s) =\frac{\va(s)+M}{2M}.
\]
Hence it follows that for rationals $\alpha,l,$ and $u$, such that $l \leq u$, we have
\[
\va(s) > \alpha \text{ iff } \va^+(s) > \frac{\alpha +M}{2M}; \quad \text{and}
\quad \va^+(s) \in [l,u] \text{ iff } \va(s) \in [M \cdot (2l - 1) , M \cdot(2 u - 1)].
\]
Given a rational $\vare>0$, to obtain an interval $[l_1,u_1]$ such that
$u_1-l_1 \leq \vare$ and $\va(s) \in [l_1,u_1]$, we first obtain
an interval $[l,u]$ such that $u-l \leq \frac{\vare}{2M}$ and
$\va^+(s) \in [l,u]$.
From the interval $[l,u]$  we obtain the interval
$[l_1,u_1]=[ M\cdot(2l -1), M\cdot(2u -1)]$ such that $\va(s) \in [l_1,u_1]$
and $u_1 -l_1= 2\cdot M \cdot (u -l) \leq \vare$.
Hence we present the algorithm to approximate the values for normalized
rational stochastic games.

\begin{algorithm}[t]
\caption{Approximating the value of a stochastic game}
\label{algo:conc-lim-avg}
{
\begin{tabbing}
aa \= aa \= aa \= aa \= aa \= aa \= aa \= aa  \= aa \= aa \kill
\> \\
\> {\bf Input:} a normalized rational stochastic limit-average game $G$, \\
\>\>\> a state $s$ of $G$, and a rational value $\vare>0$ specifying the desired tolerance.  \\
\> {\bf Output:} a rational interval $[l,u]$ such that $u-l \leq 2 \vare$ and $\va(s) \in [l,u]$.\\
\\
\> 1. $l:=0; \ u:=1; \ m=\frac{1}{2}$; \\
\> 2. {\bf repeat}  for $\lceil\log\big(\frac{1}{\vare}\big) \rceil$ steps \\
\>\> 2.1. {\bf if} $\Phi(s,m) $, {\bf then}  \\
\>\>\> 2.1.1. $l:= m;\  u:=u; \ m:=\frac{l+u}{2}$; \\
\>\> 2.2. {\bf else} \\
\>\>\> 2.2.1. $l:=l; \ u:=m;\ m:=\frac{l+u}{2}$; \\
\> 3. {\bf return} $[l,u]$; \\
\end{tabbing}
}
\end{algorithm}

\medskip\noindent{\bf Running time of Algorithm~\ref{algo:conc-lim-avg}.}
In Algorithm~\ref{algo:conc-lim-avg} we denote by $\Phi(s,m)$ the sentence to
specify that $\va(s) > m$, and by Theorem~\ref{thrm:complexity} the
truth of $\Phi(s,m)$ can be decided in time
\[
2^{O\big(|\trans| \cdot \log(|\trans|)\big)} \cdot O(|G|^2 + |m|^2),
\]
for a stochastic game $G$, where $|m|$ is the number of bits required to
specify $m$.
In Algorithm~\ref{algo:conc-lim-avg}, the variables $l$ and $u$ are initially set
to $0$ and $1$, respectively.
Since the game is normalized,
the initial values of $l$ and $u$ clearly provide lower and upper bounds
on the value, and provide starting bounds for the binary search.
In each iteration of the algorithm, in Steps 2.1.1
and 2.2.1, there is a division by $2$.
It follows that after $i$ iterations $l,u$, and $m$ can be expressed as
$\frac{q}{2^i}$, where $q$ is an integer and $q \leq 2^i$.
Hence $l,u$, and $m$ can always be expressed in
\[
O\big(\log\big(\frac{1}{\vare}\big)\big)
\]
bits.
The loop in Step~4 runs for
$\lceil\log\big(\frac{1}{\vare}\big)\rceil=O\big(\log\big(\frac{1}{\vare}\big)\big)$
iterations, and every iteration can be computed in time
$2^{O\big(|\trans|\cdot \log(|\trans|)\big)} \cdot
O\big(|G|^2 + \log^2\big(\frac{1}{\vare}\big)\big)$.
This gives the following theorem.

\begin{theorem}{}\label{thrm:algorithm}
Given a normalized rational stochastic limit-average game $G$,
a state $s$ of $G$, and a rational $\vare>0$,
Algorithm~\ref{algo:conc-lim-avg} computes an interval $[l,u]$ such that
$\va(s) \in [l,u]$ and $u -l \leq 2 \vare$,
in time
\[
2^{O\big(|\trans| \cdot \log(|\trans|)\big)} \cdot
O\bigg(|G|^2 \cdot \log\big(\frac{1}{\vare}\big) + \log^3\big(\frac{1}{\vare}\big)\bigg).
\]
\end{theorem}

The reduction from rational stochastic games to normalized stochastic games
suggest that for a rational stochastic game $G$ and a rational tolerance
$\vare>0$,
to obtain an interval of length at most $\vare$ that contains the value
$\va(s)$, it suffices to obtain an interval of length of at most
$\frac{\vare}{2M}$ that contains the value in the corresponding normalized game,
where $M=\max\set{\abs(r(s,a,b))
\mid s \in S, a \in \mov_1(s), b \in \mov_2(s)}$.
Since $M$ can be expressed in $|G|$ bits, it follows that the size of
the normalized game is $O(|G|^2)$.
Given a tolerance $\vare>0$ for the rational stochastic game,
we need to consider the tolerance $\frac{\vare}{2\cdot M}$ for the normalized game.
The above analysis along with Theorem~\ref{thrm:algorithm} yields the
following corollary (the corollary is obtained from
Theorem~\ref{thrm:algorithm} by substituting $|G|$ by $|G|^2$, and
$\log\big(\frac{1}{\vare}\big)$ by
$|G|\cdot \log\big(\frac{1}{\vare}\big)$).

\begin{corollary}\label{coro:running-time-conclimavg}
Given a rational stochastic limit-average game $G$,
a state $s$ of $G$, and a rational $\vare>0$, an interval $[l,u]$ such that
$\va(s) \in [l,u]$ and $u -l \leq 2 \vare$,
can be computed in time
\[
2^{O\big(|\trans| \cdot \log(|\trans|)\big)} \cdot
O\bigg(|G|^5 \cdot \log\big(\frac{1}{\vare}\big) + |G|^3 \cdot\log^3\big(\frac{1}{\vare}\big)\bigg).
\]
\end{corollary}

\medskip\noindent{\bf The complexity class EXPTIME.}
A problem is in the complexity class EXPTIME~\cite{Papa94} if there is an algorithm
${\cal A}$ that solves the problem, and there is a polynomial $p(\cdot)$
such that for all inputs $I$ of $|I|$ bits, the running time of the algorithm
${\cal A}$ on input $I$ can be bounded by $2^{O(p(|I|))}$.
In case of rational stochastic limit-average games, the input is the size of the game $G$, i.e.,
the input requires $|G|$ bits.
Hence from Theorem~\ref{thrm:complexity} and Corollary~\ref{coro:running-time-conclimavg}
we obtain the following result.

\begin{theorem}
Given a rational stochastic limit-average game $G$,
a state $s$ of $G$,
rational $\vare>0$, and  rational $\alpha$, the following assertions hold.
\begin{enumerate}
\item \emph{(Decision problem)} Whether $\va(s) > \alpha$ can be decided in EXPTIME.
\item \emph{(Approximation problem)} An interval $[l,u]$ such that $u-l \leq 2\vare$ and
$\va(s) \in [l,u]$ can be computed in EXPTIME.
\end{enumerate}
\end{theorem}

\newcommand{\distance}{\mathrm{\rho}}
\medskip\noindent{\bf Approximate analysis of games with approximate description.}
Let $G=\langle S,\moves, \mov_1, \mov_2, \trans, r \rangle$ and 
$G'=\langle S,\moves, \mov_1, \mov_2, \trans', r' \rangle$
be two stochastic games such that
\begin{enumerate}
\item for all $s,t \in S$ and for all $a \in \mov_1(s)$ and $b \in \mov_2(s)$,
	we have 
	\[
	\trans(s,a,b) < (1+\eta) \cdot \trans'(s,a,b)(t) \quad \text{and} \quad
	\trans'(s,a,b) < (1+\eta) \cdot \trans(s,a,b)(t),
	\]
	for $\eta < \frac{1}{2|S|}$; and
\item for all $s \in S$ and for all $a \in \mov_1(s)$ and all $b \in \mov_2(s)$ 
	we have 
	\[
	\abs(r(s,a,b)-r'(s,a,b)) \leq \gamma.
	\]
\end{enumerate}
Let $\distance(G,G')$ be defined as the infimum over 
$\big(\frac{2\eta \cdot |S|}{(1-2\eta\cdot |S|)}\cdot ||r|| + \gamma\big)$,
where $\eta,\gamma$ ranges over all pairs that satisfy the above two inequalities.
From the result of~\cite{Sol03} it follows that the absolute difference
in the values of a player at all states in $G$ and $G'$ is bounded by
$\distance(G,G')$.
Hence given a game $G$ and an auxiliary game $G'$ that approximates 
$G$ within $\eta$, i.e., $\distance(G,G') \leq \eta$, we can 
approximate the values of the game $G'$ for $\vare>0$, and obtain a 
$\eta+\vare$ approximation of the values of the game $G$.
This enables us to approximate the values of stochastic games described
approximately.

Unfortunately, the only lower bound we know on the complexity of the
decision problem is PTIME-hardness (polynomial-time hardness).
The hardness follows from a reduction from alternating reachability~\cite{Bee80,Immerman81}.
Even for the simpler case of perfect-information deterministic games, no
polynomial time algorithm is known~\cite{ZP95}, and
the best known deterministic algorithm for perfect information games
is exponential in the size of the game.
In case of perfect-information stochastic games, deterministic and stationary
optimal strategies exist~\cite{LiggettLippman69}.
Since the number of deterministic stationary strategies can be at most
exponential  in the size of the game, there is an exponential time algorithm
to compute the values exactly (not approximately)
(also see the survey~\cite{NeySor:Book}).
From the polynomial time algorithm to compute values in Markov decision processes
\cite{FV97} and the existence of pure stationary optimal strategies in
perfect-information games~\cite{LiggettLippman69}, it follows that the
decision problem for perfect-information games lie in NP $\cap$ coNP.
Better complexity bounds than EXPTIME to solve the decision and the
approximation problem for stochastic games is an interesting
open problem; and even for deterministic games no better bound is known.

\subsection*{Acknowledgments}
We are grateful to Prof. Abraham Neyman for his comments on an earlier draft of
the paper. His comments helped us in improving the presentation of the paper
vastly and in making the proofs more precise.
We thank an anonymous referee for useful comments.

\end{document}